\documentclass[12pt]{article}
\usepackage{epsfig, epsf, graphicx, subfigure}
\usepackage{pstricks, pst-node, psfrag}
\usepackage{amssymb,amsmath,bm}
\usepackage{verbatim,enumerate}
\usepackage{rotating, lscape}
\usepackage{setspace}
\usepackage[english]{babel}
\usepackage[sort&compress]{natbib}
\usepackage{multirow}
\usepackage{theorem}

\def\Cor{\mathrm{cor}}

\def\d{\mathrm{d}}
\def\Var{\mathrm{var}}

\def\p{\partial}

\def\rrho{\boldsymbol{\rho}}
\def\aalpha{\boldsymbol{\alpha}}
\def\Ssigma{\boldsymbol{\Sigma}}
\def\00{\mathrm{0}}

\def\ZZ{\mathbf{Z}}

\def\ss{\mathbf{s}}
\def\TT{\mathrm{T}}
\def\EE{\mathrm{E}}
\def\WW{\mathbf{W}}
\def\tht{\boldsymbol{\theta}}

\def\pr{\mathrm{pr}}

\newtheorem{prop}{Proposition}

\begin{document}

\thispagestyle{empty} \baselineskip=28pt \vskip 5mm
\begin{center} {\Huge{\bf Factor Copula Models for Replicated Spatial Data}}
\end{center}

\baselineskip=12pt \vskip 10mm

\begin{center}\large
Pavel Krupskii, Rapha\"{e}l Huser, Marc G. Genton\footnote[1]{
\baselineskip=10pt CEMSE Division,
King Abdullah University of Science and Technology,
Thuwal 23955-6900, Saudi Arabia. E-mail: pavel.krupskiy@kaust.edu.sa, raphael.huser@kaust.edu.sa, marc.genton@kaust.edu.sa\\
This research was supported by the
King Abdullah University of Science and Technology (KAUST).}
\end{center}

\baselineskip=17pt \vskip 10mm \centerline{\today} \vskip 15mm

\begin{center}
{\large{\bf Abstract}}
\end{center}

We propose a new copula model that can be used with replicated spatial data. Unlike the multivariate normal copula, the proposed copula is based on the assumption that a common factor exists and affects the joint dependence of all measurements of the process. Moreover, the proposed copula can model tail dependence and tail asymmetry. The model is parameterized in terms of a covariance function that may be chosen from the many models proposed in the literature, such as the Mat\'{e}rn model. For some choice of common factors, the joint copula density is given in closed form and therefore likelihood estimation is very fast. In the general case, one-dimensional numerical integration is needed to calculate the likelihood, but estimation is still reasonably fast even with large data sets. We use simulation studies to show the wide range of dependence structures that can be generated by the proposed model with different choices of common factors. We apply the proposed model to spatial temperature data and compare its performance with some popular geostatistics models.

\baselineskip=14pt

\par\vfill\noindent
{\bf Some key words:}
copula; heavy tails; non-Gaussian random field; spatial statistics; tail asymmetry.
\par\medskip\noindent
{\bf Short title}: Factor Copula Models for Replicated Spatial Data

\clearpage\pagebreak\newpage \pagenumbering{arabic}
\baselineskip=26pt

\section{Introduction}

Flexible but simple and interpretable models are often needed to model spatial data. Models based on multivariate normality have been widely used for modeling spatial data and for interpolation at new locations, as well as for uncertainty analysis. However, Gaussian models do not account for strong dependencies in the tails and asymmetric dependencies between left and right tails, which are often found in real data. More flexible
models that retain the appealing tractability of Gaussian random fields are therefore needed.

To model data with asymmetric dependencies and strong tail dependence, a copula-based approach is particularly convenient.
Copula models have been applied in a wide range of actuarial, financial and environmental studies; see \citet{Krupskii.Joe2015b}, \citet{Genest.Favre2007}, \citet{Patton2006}, \citet{Jondeau.Rockinger2006} among others. A copula is defined as a multivariate cumulative distribution function with uniform $U(0,1)$ margins; it may be used to link univariate marginals to construct a joint distribution. For a continuous $n$-dimensional cumulative distribution function, $F$, with univariate margins, $F_1,\ldots,F_n$, \citet{Sklar1959} showed that a unique copula, $C$, exists such that $F(x_1,\ldots,x_n) = C\{F_1(x_1),\ldots,F_n(x_n)\}$. In practice, inference is typically performed in two steps: univariate marginal distributions, $F_1,\ldots,F_n$, are first estimated and a copula is then used to model the joint dependencies governing the data transformed to the uniform scale. To increase efficiency of the estimates, both marginal and copula parameters can be estimated jointly in one step. A detailed overview of copulas is presented by \citet{Nelsen2006} and \citet{Joe2014}.

Many copulas proposed in the literature, however, are not suitable for modeling spatial data. For example, multivariate Archimedean copulas have
exchangeable dependence structures, while spatial processes typically have stronger dependencies at smaller distances. The factor copula models proposed by \citet{Krupskii.Joe2015b} are not permutation symmetric, although the order of variables is not important when dependencies between different observations from a spatial process are modeled. The Farlie-Gumbel-Morgenstern copula studied by \citet{Farlie1960}, \citet{Gumbel1960b} and \citet{Morgenstern1956} is permutation symmetric but this model is suitable only for modeling weak dependencies between variables.

The Gaussian copula is quite convenient for modeling spatial data as it may be parameterized in terms of a covariance function that
controls the strength of dependencies as a function of distance. Usually, covariance functions impose a monotonically decaying correlation with distance, which is often realistic in applications. Furthermore, Gaussian conditional distributions are available in closed form and Gaussian data are easily simulated.
However, the Gaussian copula lacks tail dependence and is reflection symmetric. The Student's $t$ copula can handle tail dependence, but is reflection symmetric, similarly to the Gaussian copula. The skew-$t$ and skew-normal copulas obtained from skew-$t$ and skew-normal distributions, respectively, may not be suitable as the relationship between asymmetry and tail dependence is quite complicated and quantile calculations can be computationally demanding; see for example the skew-$t$ distribution of \citet{Azzalini.Capitanio2003}. Extreme-value copulas are tailored for extremes and can capture tail dependence and asymmetry \citep{Segers.2012}, but computation of joint densities is excessively prohibitive in high dimensions \citep{Castruccio.Huser.ea2016}, which makes them difficult to use. More importantly, they are justified for the extremes but may not be suitable for data in the center of the distribution. Figure 1 in \citet{Li.Genton.2013} depicts the relationships among various copula structures.

Alternatively, one may use vine copula models, the joint distribution of which is constructed using bivariate, conditional linking copulas, including models with general structures, R-vines, and special cases such as C-vines; see \citet{Kurowicka.Cooke2006} and \citet{Aas.Czado.ea2009} for details. C-vines may be used to model joint dependencies and $k=5$--$10$ closest neighbors may be used for interpolation with the help of the C-vine copula rooted in an unknown location. To achieve greater flexibility, different copula families and convex combinations may be used; see \citet{Graler.Pebesma2011} and \citet{Graler2014}. More general R-vine models may be selected and different spatial covariates can be included to reduce the number of dependence parameters in the model; see \citet{Erhardt.Czado.ea2014}. However, these models lack interpretability and their dependence structures depend on the likelihood value. Also, estimation for high-dimensional data can be very time consuming.

For spatial data, it is natural to have a parameterization in terms of pairwise dependencies. \citet{Bardossy.Li2008} advocated a V-transformed copula obtained from a non-monotonic transformation of multivariate normal variables and \citet{Bardossy2011} used their copula for modeling and interpolation of asymmetric groundwater data. The chi-squared asymmetric copula of \citet{Bardossy2006} was obtained similarly. The main drawback of such marginally transformed normal variables is that the resulting copula is tail independent and the likelihood is a sum of $2^n$ terms, where $n$ is the number of locations. Parameter estimation and interpolation using these models is thus not an easy task.

Spatial factor models, which are based on the assumption that a common latent random factor
affects all spatial locations simultaneously, turn out to remedy many of the drawbacks described above. These models include the generalized common factor spatial models of \citet{Wang.Wall2003}, \citet{Hogan.Tchernis2004} and \citet{Irincheeva.Cantoni.ea2012}, the nonparametric model of \citet{Christensen.Amemiya2002} and others. Factor models are appealing as they may be 
interpreted in many applications in which an unobserved variable explains the dependence between measured variables. However, no flexible copulas associated with these models have been discussed in the literature, and the dependence properties of these models have not been studied in detail.

We propose a model that combines the flexibility of a copula modeling approach, the interpretability and parsimony of factor models, the tractability of the Gaussian copula in high dimensions, and that may be efficiently fitted to spatial data with temporal replicates. The model and the corresponding copula are based on the following random process:
\begin{equation} \label{base_model}
W(\ss) = Z(\ss) + V_0, \quad \ss \in \mathbb{R}^d,
\end{equation}
where $Z(\ss)$ is a Gaussian process and $V_0$ is a common factor, which does not depend on the spatial location, $\ss$. A skew-Gaussian random field is the special case of this model when $V_0 = |Z_0|$, $Z_0 \sim N(0,1)$. \citet{Genton.Zhang2012} discussed some identifiability issues with this model when applied to purely spatial data (i.e., with no replicates) and proposed some simple remedies. However, their approach is not applicable to the more general case in (\ref{base_model}); replicates are needed to estimate parameters in (\ref{base_model}). With an appropriate choice of the common factor, $V_0$, model (\ref{base_model}) allows for both tail dependence and asymmetric dependence between the two tails, and is parameterized in a way that is convenient for spatial data. Parameter estimation may be efficiently performed using an adjusted Newton-Raphson algorithm even if the data dimension is fairly large.

The rest of the paper is organized as follows. In Section \ref{sec_factmodel} we study model (\ref{base_model}) and its tail properties, and illustrate some examples with different choices of common factor $V_0$. In Section \ref{sec_MLE}, we give some details on likelihood estimation and derive the conditional distributions that may be used for spatial interpolation (i.e., kriging). We conduct a simulation study in Section \ref{sec_empstudy} to show the performance of the estimation procedure. 
We also apply the proposed model to temperature data and compare its performance with some classical models used in geostatistics. Section \ref{sec_discussion} concludes with a discussion. 

\section{The Common Factor Model for Spatial Data}
\label{sec_factmodel}

\subsection{Model and tail properties}
\label{sec_factmodel_sub0}

We use the following notation:
$\Phi(\cdot)$ is the cumulative distribution function of the univariate standard normal variable, whereas $\Phi_{\Ssigma}(\cdot)$ is that of a multivariate standard normal random variable with correlation matrix $\Ssigma$. Small symbols denote the corresponding densities.

We consider measurements of a random process in a specific area that is not very large, or, at least, is quite homogeneous. We assume that there exists an unobserved random factor that affects the joint dependence of all measurements of this process. Specifically, we construct the corresponding copula by restricting  model (\ref{base_model}) to a finite set of locations $\ss_1,\ldots,\ss_n \in \mathbb{R}^d$. By simplicity, we write $Z_j = Z(\ss_j), W_j = W(\ss_j)$, $j=1,\ldots,n$, and $\ZZ=(Z_1,\ldots,Z_n)^{\TT}$, $\WW = (W_1,\ldots,W_n)^{\TT}$. The finite-dimensional restricted model is
\begin{equation}
\label{mainmodel}
W_j = Z_j + V_0, \quad j=1,\ldots,n, \quad \ZZ \sim N(\mathbf{0}, \Ssigma_{\ZZ}),
\end{equation}
where $\Ssigma_{\ZZ}$ is a correlation matrix and $V_0 \sim F_{V_0}$ is a common factor that is independent of $\ZZ$. 
It can be verified that the correlation matrix of $\WW$ is
$$\Ssigma_{\WW} := \Cor(\WW) = (\Ssigma_{\ZZ} + \sigma^2_0)/(1 + \sigma^2_0),$$
assuming that the variance $\sigma^2_0 =\Var(V_0)$ exists.
The correlation $\Ssigma_{\WW,j_1,j_2}$ decreases as the correlation $\Ssigma_{\ZZ,j_1,j_2}$ decreases. Moreover, the matrix $\Ssigma_{\ZZ}$ can be parameterized using a correlation function for spatial data as, e.g., the exponential or Mat\'{e}rn correlation function.

Note that $\Ssigma_{\ZZ,j_1,j_2} = 1$ is equivalent to $\Ssigma_{\WW,j_1,j_2} = 1$, but $\Ssigma_{\ZZ,j_1,j_2} = 0$ implies $\Ssigma_{\WW,j_1,j_2} = \sigma^2_0/(1 + \sigma^2_0) > 0$. This is expected as model (\ref{base_model}) is in fact a mixture of $Z(\ss)$ with a perfectly dependent spatial process, $V_0$. As we will see below, this subtle construction allows us to obtain a tail dependent process in some cases. To capture low correlations, we might need to use correlation functions for $Z$ that can take negative values, such as the damped cosine function $\rho(h) = \cos(h)\exp(-\lambda h), \lambda > 0$; see \citet{Gneiting.Genton.ea2007} for a review of correlation functions. Although this is an oscillating correlation function and such a behavior is rarely seen in practice, it can be appropriate for modeling data in a small domain when the maximum distance $H$ is not very large and $\rho(h)$ is a decreasing function for $h \in (0,H)$. Alternatively, one could assume 
that distinct independent (or weakly dependent) random factors affect different regions separately. This could be a possible extension of the proposed model for handling data in a larger domain with nearly independent observations at large distances. In this paper we focus on modeling spatial data in a small domain and hence assume model (\ref{base_model}). 
All realizations from this random process will be dependent; however, the dependence is weaker for pairs separated by larger distances.

We now obtain the distribution of the vector $\WW$ in (\ref{mainmodel}) in a general form. We have:
$$
F_{n}^{\WW}(w_1,\ldots,w_n) 
= \int_{-\infty}^{\infty}\Phi_{\Ssigma_{\ZZ}}(w_1-v_0,\ldots,w_n-v_0)\d F_{V_0}(v_0),
$$
and the density may therefore be expressed in terms of a one-dimensional integral; that is,
$$
f_{n}^{\WW}(w_1,\ldots,w_n) = \int_{-\infty}^{\infty}\phi_{\Ssigma_{\ZZ}}(w_1-v_0,\ldots,w_n-v_0)\d F_{V_0}(v_0).
$$

Consequently, the resulting copula and its density may be expressed as
{\small
$$
C_{n}^{\WW}(u_1,\ldots,u_n) = F_{n}^{\WW}\left\{(F_1^{\WW})^{-1}(u_1),\ldots,(F_1^{\WW})^{-1}(u_n)\right\}, \quad F_1^{\WW}(w_1) = \int_{-\infty}^{\infty}\Phi(w_1-v_0)\d F_{V_0}(v_0),
$$
\begin{equation}
c_{n}^{\WW}(u_1,\ldots,u_n) = \frac{f_{n}^{\WW}\left\{(F_1^{\WW})^{-1}(u_1),\ldots,(F_1^{\WW})^{-1}(u_n)\right\}}{f_1^{\WW}\left\{(F_1^{\WW})^{-1}(u_1)\right\}\times\cdots\times f_1^{\WW}\left\{(F_1^{\WW})^{-1}(u_n)\right\}}. \label{cop_pdf}
\end{equation}}

Note that the distribution of $W_1$, $F_1^{\WW}$, is only used to construct the copula $C_{n}^{\WW}$ which, in turn, can be used to model the joint dependence for data with any univariate marginals.

The distribution of the common factor, $V_0$, in (\ref{mainmodel}) determines the tail properties of the resulting copula. In spatial applications, data often show stronger dependencies in the tails than is predicted using a multivariate normal copula. So-called tail dependence coefficients are standard measures of tail dependence for a pair of variables, used in the copula literature. For any bivariate copula, $C$, such coefficients are defined as limiting quantities:
\begin{equation}
\label{tail_dep_coef}
\lambda_L := \underset{q\to 0}{\lim} \,C(q, q)/q \in [0, 1]
\quad \text{ and } \quad \lambda_U := \underset{q\to 0}{\lim} \,\bar C(1-q, 1-q)/q \in [0, 1],
\end{equation}
where $\bar C(u_1,u_2) := 1-u_1-u_2+C(u_1,u_2)$ is the survival copula. The copula, $C$, has lower or upper tail dependence if $\lambda_L > 0$ or $\lambda_U > 0$, respectively. For $(U_1,U_2) \sim C$, $\lambda_L = \lim_{q\to 0} \pr(U_1 \leq q| U_2 \leq q)$ and $\lambda_U = \lim_{q\to 1} \pr(U_1 \geq q| U_2 \geq q)$. For copulas with tail dependence, the limiting conditional probabilities of extreme events are therefore positive. For the normal copula, $\lambda_L = \lambda_U = 0$. This means that a standard model based on multivariate normality might underestimate the joint probability of extreme events
and their probability of simultaneous occurrence. Asymmetric tail dependence with $\lambda_L \neq \lambda_U$ is often found in data as well, a feature that the Gaussian copula cannot capture.

If the common factor $V_0$ in (\ref{mainmodel}) is Gaussian, the joint distribution of $\WW$ is multivariate normal and therefore there is no modeling gain. Furthermore, the resulting model is overparameterized. To generate tail dependence, we need to use a distribution for $V_0$ that has heavier tails than the normal distribution. The normal density has a quadratic exponential order of tail decay, and tail dependence can be obtained using a random variable, $V_0$, with a sublinear exponential order of decay, as Proposition \ref{prop-1} shows. The proof is in the Appendix. 
\begin{prop}  \label{prop-1} \rm
Let $1-F_{V_0}(v_0) \sim Kv_0^{\beta}\exp(-\theta v_0^{\alpha})$, $\alpha \geq 0$, $\beta \in \mathbb{R}$, $\theta > 0$, $K > 0$, as $v_0 \to \infty$. Let $\rho = \Ssigma_{\ZZ,1,2} < 1$. If $0 < \alpha < 1$ or $\alpha = 0, \beta < 0$, then the bivariate copula, $C_{2}^{\WW}(u_1,u_2)$, has perfect upper tail dependence, i.e., $\lambda_U=1$. If $\alpha = 1$, the copula $C_{2}^{\WW}(u_1,u_2)$ has upper tail dependence with $\lambda_U=2\Phi\left[-\theta\{(1-\rho)/2\}^{1/2}\right]$. If $\alpha > 1$, the copula $C_{2}^{\WW}(u_1,u_2)$ is tail independent, i.e., $\lambda_U = 0$. Similar results hold for the lower tail.
\end{prop}

Following Proposition \ref{prop-1}, tail dependence can also be obtained when using a common factor that has power law decay. The examples of such distributions include the Student's $t$ distribution, the Pareto distribution and others. We now list some interesting models.

\emph{Example 1:} Consider the Weibull distribution, $F_{V_0}(v_0)  = 1 - \exp(-\theta v_0^{\alpha})$, $v_0 > 0$ and $\theta > 0,\, \alpha > 0$. It is easy to check that the marginal distribution may be expressed as $F_1^{\WW}(w) = \Phi(w) - (2\pi)^{-1/2}\int_{-\infty}^w\exp\{-\theta(w-z)^{\alpha}-z^2/2\}\d z$.
When $\alpha = 1$, i.e., $V_0$ is exponentially distributed, the formula simplifies to $F_1^{\WW}(w) = \Phi(w) - \exp(\theta_0^2/2-\theta_0w)\Phi(w-\theta_0)$, which can be easily numerically inverted and used in (\ref{cop_pdf}) to calculate the likelihood. In the general case with $\alpha \neq 1$, the computation of the inverse distribution function, $(F_1^{\WW})^{-1}$, is still relatively easy but requires more time. Moreover, it can be seen that, for $\alpha \leq 1$, the resulting bivariate copula, $C_{2}^{\WW}$, is asymmetric with upper tail dependence: if $\alpha = 1$, $\lambda_U = 2\Phi\left[-\theta\{(1-\rho)/2\}^{1/2}\right]$, and if $0 < \alpha < 1$, $\lambda_U = 1$ (perfect co-monotonic tail dependence). By contrast, if $\alpha > 1$, the copula, $C_{2}^{\WW}$, is upper tail independent. In other words, tail properties of this copula depend on the shape parameter, $\alpha$, of the Weibull factor.

\emph{Example 2:} Let $V_0 = V_1 - V_2$, where $V_1 \sim \mathrm{Exp}(\theta_1)$, $V_2 \sim \mathrm{Exp}(\theta_2)$ are independent exponential random factors with distribution $F_{V_j}(v_0) = 1 - \exp(-\theta_j v_0)$, $v_0 > 0$ and $\theta_j > 0$ ($j=1,2$). We can find that $F_{V_0}(v_0) = \exp\{-\theta_2{(-v_0)}_+\}\left\{1-\theta_2\exp(-\theta_1 {v_0}_+)/(\theta_1+\theta_2) \right\}$, where ${v_0}_+ = \max\{v_0,0\}.$
We can then obtain a simple formula for the marginal distribution: $F_1^{\WW}(w) = \Phi(w) - \exp(\theta_1^2/2-\theta_1w)\Phi(w-\theta_1)\theta_2/(\theta_1+\theta_2) + \exp(\theta_2^2/2+\theta_2w)\Phi(-w-\theta_2)\theta_1/(\theta_1+\theta_2).$
Consequently, the inverse distribution function can be quickly computed using numerical inversion. Furthermore, the copula density can be computed in closed form. 
From Proposition \ref{prop-1}, the resulting copula, $C_{2}^{\WW}$, is asymmetric whenever $\theta_1 \neq \theta_2$ with lower and upper tail dependence coefficients $\lambda_L = 2\Phi\left[-\theta_2\{(1-\rho)/2\}^{1/2}\right]$, $\lambda_U = 2\Phi\left[-\theta_1\{(1-\rho)/2\}^{1/2}\right]$.

\emph{Example 3:} Consider the Pareto distribution with $F_{V_0}(v_0) = 1 - (v_0/\theta)^{-\beta}$, $v_0 > \theta$ and $\theta > 0$, $\beta > 0$. In practice it may be useful to restrict $\beta > 2$, so that the variance of $V_0$ exists and we can obtain decreasing correlations with larger distances. We find that $F_1^{\WW}(w) = \Phi(w-\theta) - \theta^{\beta}(2\pi)^{-1/2}\int_{-\infty}^{w-\theta}(w-z)^{-\beta}\exp(-z^2/2)\d z$. From Proposition \ref{prop-1}, the resulting copula, $C_{2}^{\WW}$, is asymmetric with perfect (co-monotonic) upper tail dependence, i.e., $\lambda_U = 1$.

\emph{Remark 1}. When the copula $C_2^{\WW}$ has $\lambda_U = 0$ or $\lambda_U = 1$, the limiting extreme value copula is the independence or co-monotonic copula, respectively; see, for example, \citet{Joe.Li.ea2010}, \citet{Hua.Joe2011, Hua.Joe2012}. As a result, the limiting copula in the Weibull factor model with $\alpha > 1$ (Example 1) will be the independence copula. By contrast, for the Weibull factor model with $\alpha < 1$ and for the Pareto factor model (Example~3), the co-monotonic copula arises as the limiting extreme value copula. We next show that the stable upper tail dependence function \citep{Segers.2012} of the limiting extreme-value copula in the exponential factor model (Example 2) corresponds to the \cite{Husler.Reiss1989} distribution, which has been widely used as a flexible family to model extreme events \citep{Davison.Huser.ea2013, Davison.Padoan.ea2012, Huser.Davison2013, Thibaud.Mutzner.ea2013}. The proof of this result is in the Supplementary Material. 
A similar result holds for the lower tail.

\begin{prop} \label{prop-2} \rm
Let $\ell_q(x_1,x_2) = \{1-C_{2}^{\WW}(1-qx_1,1-qx_2)\}/q$. Then in the exponential factor model (Example 2) we get $\lim_{q\to 0}\ell_q(x_1,x_2) = x_1\Phi\left\{\lambda/2 + \log(x_1/x_2)/\lambda\right\} + x_2\Phi\left\{\lambda/2 + \log(x_2/x_1)/\lambda\right\}$ where $\lambda = \theta_1\{2(1-\rho)\}^{1/2}$.
\end{prop}
This result describes the type of the joint distribution that arises at high levels (i.e., in the joint upper tail), or equivalently the limiting distribution for normalized componentwise maxima of i.i.d. copies of the process observed at two sites. One can generalize this result and show that the process described by model (\ref{base_model}) is in the max-domain of attraction of the (max-stable) Brown-Resnick process \citep{Kabluchko.ea2009}. However, although the asymptotic tail properties of our model and the Brown-Resnick model are similar, inference is much easier using our model; see Section \ref{sec_MLE} and \cite{Castruccio.Huser.ea2016}.

\subsection{Tail flexibility for different factors}
\label{sec_factmodel_sub1}

To show the flexibility of our model, we assume that the common factor in (\ref{mainmodel}) can be decomposed as $V_0 = V_1 - V_2$, where $V_1 \geq 0$ and $V_2 \geq 0$ are independent random variables controlling the strength of the joint dependence in the upper and lower tails, respectively. We consider bivariate data with $W_j = Z_j  + V_0$ ($j=1,2$), where $(Z_1, Z_2)$ has a bivariate standard normal distribution with correlation parameter $\rho_Z := \Cor(Z_1, Z_2)$. We select three models and calculate $\lambda_L^q := C_2^{\WW}(q,q)/q, \, \lambda_U^q := [2q-1+C_2^{\WW}(1-q,1-q)]/q$ for different values of $q$, where $\lambda_L^q$ and $\lambda_U^q$ converge as $q \to 0$ to the lower and upper tail dependence coefficients, $\lambda_L$ and $\lambda_U$, respectively. The ratio $A(q) = \lambda_L^q/\lambda_U^q$, $0 \leq q \leq 0.5$, can be used as a measure of asymmetry. If $C_2^{\WW}$ is a symmetric copula, $A(q) \equiv 1$. If $A(q) >1$ ($A(q) < 1$), dependence along the main diagonal is stronger in the lower (upper, respectively) tail. In addition, we compute $\zeta_1=\EE(U_1+U_2-1)^3$ for $(U_1,U_2) \sim C_2^{\WW}$. The measure $\zeta_1$ can be used as a measure of skewness for $C_2^{\WW}$; see \cite{Rosco.Joe.2013}. Unlike tail dependence coefficients, the measures of asymmetry, $A(q)$ and $\zeta_1$ cannot be obtained in closed form and therefore we rely on Monte Carlo simulations.  We focus on the following models: 1) $V_j \sim \mathrm{Exp}(\theta_j)$; 2) $V_j \sim \mathrm{Pareto}(\theta_j, \beta_j)$; 3) $V_j \sim \mathrm{Weibull}(\theta_j, \alpha_j)$ ($j=1,2$).

The distributions of the common factors are parameterized in the same way as in Examples $1$--$3$ in Section \ref{sec_factmodel}.
We select dependence parameters such that the corresponding copula, $C_2^{\WW}$, has the Spearman's $\rho$ equal to 0.3, 0.5 and 0.7 in all these models.
For the exponential common factor model (model 1) we use $\rho_Z = 0.04, \, 0.33, \, 0.60$ and $\theta_1 = 1.7,\,  \theta_2 = 3$, with stronger dependence in the upper tail and $\lambda_L = 0.04,\, 0.08,\, 0.18$ and $\lambda_U = 0.24,\, 0.33,\, 0.45$, respectively. For model 2 we use $\rho_Z = 0.08, \, 0.35, \, 0.62$ and $\theta_1 = 1.5,\,  \beta_1 = 4,\,  \theta_2 = 1,\,  \beta_2 = 5$, with stronger dependence in the upper tail. For model 3, we use $\rho_Z = 0.10, \, 0.37, \, 0.63$ and $\theta_1 = 3,\,  \alpha_1 = 0.8,\,  \theta_2 = 2.5,\,  \alpha = 0.6$, with stronger dependence in the lower tail. Models 2 and 3 have asymptotic perfect co-monotonic dependence in both tails so that $\lambda_L = \lambda_U = 1$. However, for $q > 0$ the values of $\lambda^q_L, \lambda^q_U$ are  different.

\begin{figure}
\begin{center}
\includegraphics[width=6in,height=6in]{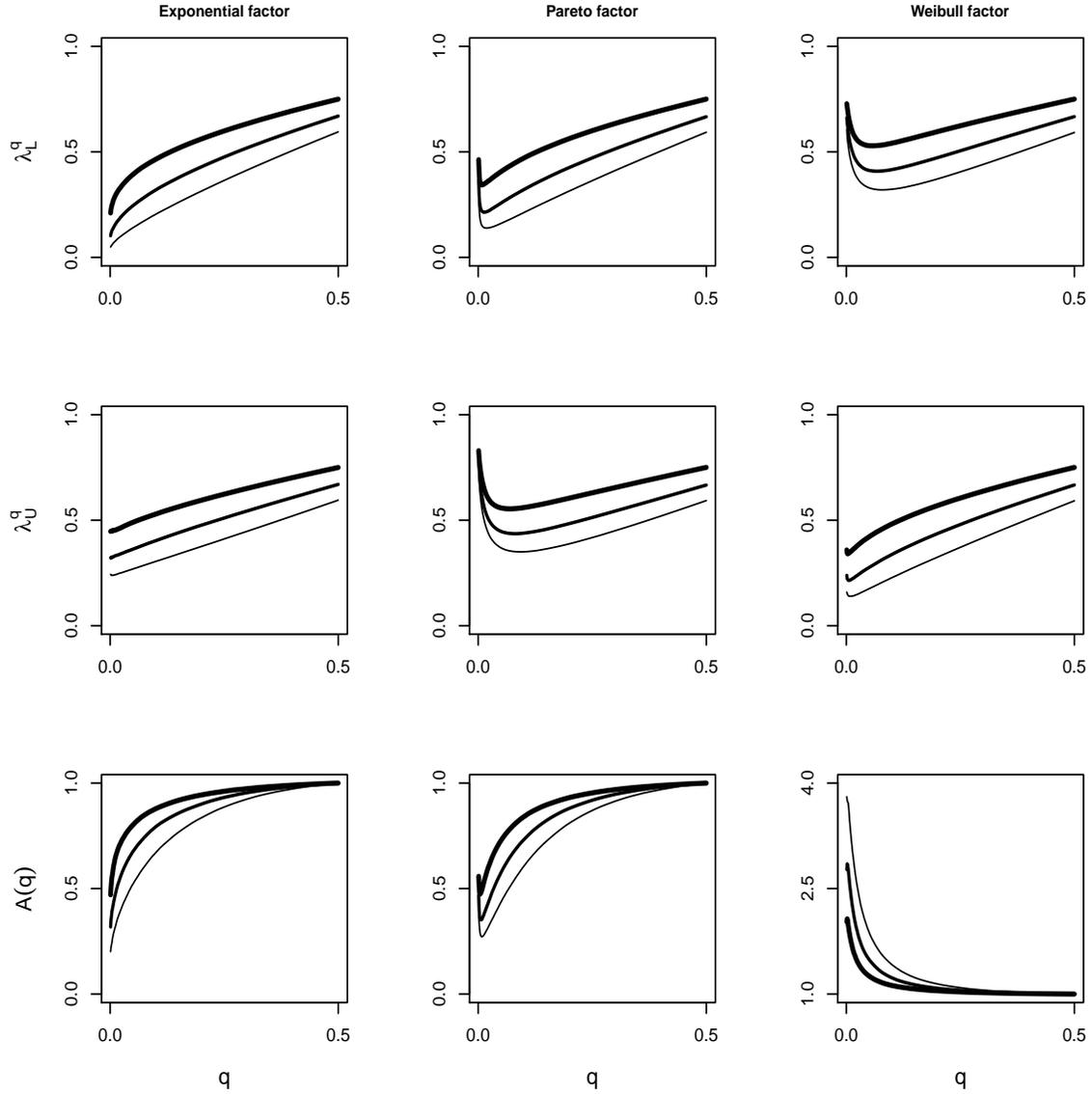}
\caption{{\footnotesize $\lambda_L^q$ (top), $\lambda_U^q$ (middle) and $A(q)$ (bottom), $0.001 \leq q \leq 0.5$, for $C_2^{\WW}$ in models 1 (left), 2 (middle) and  3 (right); Spearman's $\rho = 0.3$ (thin), $0.5$ (normal), $0.7$ (thick).}}
\label{fig_3tails}
\end{center}
\end{figure}

We plot $\lambda_L^q, \lambda_U^q$ and $A(q)$ for models 1, 2 and 3 with the Spearman's $\rho = 0.3, 0.5, 0.7$ for $0.001 \leq q \leq 0.5$ to show that we can generate models with different tail properties depending on the choice of common factors; see Fig. \ref{fig_3tails}. Tail dependence and tail asymmetry can be obtained using all these models; however, dependence in the tails is much stronger with Pareto and Weibull factors where the shape parameters, $\beta$ and $\alpha$, are smaller. Such common factors may therefore be used in applications with strong or increasing tail dependence.

We also found that $\zeta_1 = 0.007, 0.005, 0.003$ for model 1, $\zeta_1 = 0.008, 0.006, 0.004$ for model 2, and $\zeta_1 = -0.006, -0.004, -0.002$ for model 3 for Spearman's $\rho = 0.3, 0.5, 0.7$, respectively. The negative sign of $\zeta_1$ for model 3 implies stronger dependence in the lower tail unlike models 1 and 2. The maximum possible value of $|\zeta_1|$ for a continuous copula is about $0.027$ \citep{Rosco.Joe.2013}. It implies that, with stronger overall dependence, tail asymmetry (as measured by $A(q)$) and overall asymmetry (as measured by $\zeta_1$) of the copula $C_2^{\WW}$ is weaker. To capture stronger asymmetry, one therefore needs to use factors with stronger tail dependence (Pareto factors or exponential factors with larger scale parameters).

\section{Maximum Likelihood Estimation and Interpolation}
\label{sec_MLE}

\subsection{The likelihood function and its gradient}
\label{sec_MLE_sub0}

We show how to perform maximum likelihood estimation for model (\ref{mainmodel}).
We assume that we observe $N$ independent samples $\mathbf{y}_i = (y_{i1},\ldots,y_{in})^{\TT}$ ($i=1,\ldots,N$) from model (\ref{mainmodel}) with essentially arbitrary marginals, not necessarily given by the distribution function $F_1^{\WW}$. To estimate the copula parameters, we need to transform the data to a uniform scale. This can be done non-parametrically as follows: for each $j=1,\ldots,n$, we can define the uniform scores, $u_{ij} = \{\mathtt{rank}(y_{ij})-0.5\}/N$ ($i=1,\ldots,N$),
and let 
$z_{ij}(\tht_F) = (F_1^{\WW})^{-1}(u_{ij};\tht_F)$, where $\tht_F$ is a vector of parameters for $F_1^{\WW}$  ($i=1,\ldots,N$,\,  $j=1,\ldots,n$). In this case, the uniform scores are an approximation to uniform $U(0,1)$ data. From (\ref{cop_pdf}), the pseudo log-likelihood is:
\begin{equation}
\ell(\mathbf{y}_1,\ldots,\mathbf{y}_N; \tht_F, \tht_{\Ssigma}) = \sum_{i=1}^N\log f_n^{\WW}\{z_{i1}(\tht_F),\ldots,z_{in}(\tht_F);\tht_F,\tht_{\Ssigma}\} - \sum_{i,j=1}^{N,n}\log f_1^{\WW}\{z_{ij}(\tht_F);\tht_F\}, \label{loglik}
\end{equation}
where $\tht_{\Ssigma}$ is used to parameterize the correlation matrix, $\Ssigma_{\ZZ}$. 

Alternatively, one can specify parametric margins, $G_j(\cdot;\tht_j)$, for each location $j=1,\ldots,n$. The marginal parameters, $\tht_1, \ldots, \tht_n$, can be estimated in a first step and pseudo-uniform data can be obtained by applying the integral transform: $u_{ij} = G_j(y_{ij};\widehat\tht_j)$, $i=1,\ldots,N$, $j=1,\ldots,n$. Copula parameters can then be estimated in a second step using the pseudo log-likelihood in (\ref{loglik}). To increase efficiency, the marginal parameters, $\tht_G = (\tht_1^{\TT},\ldots,\tht_n^{\TT})^{\TT}$, and the copula parameters, $\tht_F, \tht_{\Ssigma}$, may be estimated jointly. Denote 
$z_{ij}(\tht_j,\tht_F) = (F_1^{\WW})^{-1}(u_{ij}(\tht_j);\tht_F)$ and $u_{ij}(\tht_j) = G_j(y_{ij};\tht_j)$, $i=1,\ldots,N$, $j=1,\ldots,n$. The full log-likelihood is \vspace{-1em}
\begin{multline}
\ell(\mathbf{y}_1,\ldots,\mathbf{y}_N; \tht_F, \tht_{\Ssigma}, \tht_G) = \sum_{i=1}^N\log f_n^{\WW}\{z_{i1}(\tht_j,\tht_F),\ldots,z_{in}(\tht_j,\tht_F);\tht_F,\tht_{\Ssigma}\}\\
 - \sum_{i=1}^N\sum_{j=1}^n\log f_1^{\WW}\{z_{ij}(\tht_j,\tht_F);\tht_F\} + \sum_{i=1}^N\sum_{j=1}^n\log g_j(y_{ij};\tht_j), \label{jloglik}
\end{multline}
where $g_j$ is the density function corresponding to $G_j$, $j=1,\ldots,n$. Here, the parameters $\tht_j = \tht(\ss_j)$ may be parameterized as a function of location $\ss_j$ and potentially other covariates to reduce the number of marginal parameters in the model.

The maximum likelihood estimates are consistent and asymptotically normal if the copula family is correctly specified. The asymptotic variance of the estimates depends on the estimation method; see \S\S 5.4, 5.5 and 5.9 of \cite{Joe2014} for the detailed review of asymptotic theory when estimates are obtained using the joint likelihood, the two-step approach with parametric margins and the two-step approach with non-parametric ranks, respectively.

For the Newton-Raphson algorithm, the first- and second-order derivatives of $\ell(\mathbf{y}_1,\ldots,\mathbf{y}_N)$ with respect to $\tht_F, \tht_{\Ssigma}$ are required at each iteration. If there is no simple form for the derivatives, they can be obtained numerically, but this requires multiple calculations of the log-likelihood. In our model, the function $\ell(\mathbf{y}_1,\ldots,\mathbf{y}_N)$ involves one-dimensional integration and therefore calculating derivatives numerically can slow down the estimation process significantly. To avoid this problem, we obtain a simple form for the first-order derivatives. We show how to calculate the gradient of the log-likelihood in the Supplementary Material. 

\emph{Remark 2}. In the special case of the exponential common factor (Example 2 in Section \ref{sec_factmodel}), 
the log-likelihood can be obtained in closed form.  With this model, numerical integration is not required to calculate the likelihood function and its derivatives. More details are given in the Supplementary Material.

\subsection{Conditional copula and interpolation} \label{sec_MLE_sub3}

In spatial applications, prediction at unobserved locations (i.e., kriging) is often required. Let $\widehat \tht_F$, $\widehat \tht_{\Ssigma}$ be estimates of $\tht_F$ and $\tht_{\Ssigma}$, respectively. For a given vector of data $(u_1,\ldots,u_n)$ on the uniform scale, observed at locations $\ss_1,\ldots,\ss_n$, we can obtain the conditional distribution at any other location $\mathbf{s_0}$:
$$\widehat C_{0|n}^{\WW}(u_0|u_1,\ldots,u_n) := \frac{\int_0^{u_0} c_{n+1}^{\WW}(\tilde u_0, u_1,\ldots,u_n;\widehat \tht_F, \widehat \tht_{\Ssigma}) \d \tilde u_0}{c_{n}^{\WW}(u_1,\ldots,u_n;\widehat \tht_F, \widehat \tht_{\Ssigma})}.$$
Using this conditional distribution, we can calculate different quantities of interest, including the conditional expectation or conditional median:
$$
\widehat m_1 := \int_0^1 \tilde u_0 \,\widehat c_{0|n}^{\WW}(\tilde u_0|u_1,\ldots,u_n)\,\d \tilde u_0, \quad \widehat q_{0.5} := (\widehat C_{0|n}^{\WW})^{-1}(0.5|u_1,\ldots,u_n),
$$
where
$$
\widehat c_{0|n}^{\WW}(\tilde u_0|u_1,\ldots,u_n) = \frac{\p \widehat C_{0|n}^{\WW}(\tilde u_0| u_1,\ldots,u_n)}{\p u_0} = \frac{c_{n+1}^{\WW}(\tilde u_0,u_1,\ldots,u_n;\widehat \tht_F, \widehat \tht_{\Ssigma})}{c_{n}^{\WW}(u_1,\ldots,u_n;\widehat \tht_F, \widehat \tht_{\Ssigma})}\,.
$$
Numerical integration can be used to compute $\widehat C_{0|n}^{\WW}(u_0|u_1,\ldots,u_n)$ and the inverse function $(\widehat C_{0|n}^{\WW})^{-1}(q|u_1,\ldots,u_n)$ can then be used for interpolation. If $\widehat G_j(\cdot) = G_j(\cdot;\widehat\tht_j)$ denotes the estimated univariate marginal distribution function at location $\ss_j$, the predicted median on the original scale is $\widehat z_{j,0.5} = \widehat G_j^{-1}(\widehat q_{0.5})$.

\section{Empirical Studies}
\label{sec_empstudy}

In this section, we check the algorithm performance using simulated data sets. We show that the exponential factor model can fit data quite well even if the distribution of the common factor is misspecified. Using a Core i5-2410M CPU@2.3 GHz, the running time to fit the exponential factor model is about 10 minutes with $n=100$ locations and $N=2000$ replicates when copula parameters are estimated using the pseudo likelihood (\ref{loglik}). The running time is about 30 minutes when both marginal and copula parameters are estimated jointly using the full likelihood (\ref{jloglik}). For the Pareto factor, the running time is longer: with $n=15$--$20$ and $N=100$--$200$, it takes about $10$--$30$ minutes using (\ref{loglik}).

\subsection{Maximum likelihood estimation for the exponential common factor model} \label{sec_empstudy_sub1b}

We performed a simulation study to check the accuracy of the maximum likelihood estimates for the exponential factor model. We calculated the maximum likelihood estimates for data sets simulated on a $3 \times 3$, $5 \times 5$ and $10 \times 10$ uniform grid on $[0,1] \times [0,1]$ (so that there are $n=9, 25, 100$ locations, respectively) from the exponential factor model with $\theta_1, \theta_2$ as the upper and lower tail parameters (see Example 2, Section \ref{sec_factmodel}), respectively. The powered-exponential correlation function, $\rho(h) = \exp(-\theta_Z h^{\alpha})$, and $N=500, 1000, 2000$ independent replicates were simulated. These experiments were repeated 500 times to calculate the bias and standard deviation for a given set of dependence parameters, $\tht = (\theta_1, \theta_2, \theta_Z, \alpha)^{\TT}$. We considered four different estimation procedures:
\begin{enumerate}
\item Known univariate margins; copula parameters are estimated using data with true $U(0,1)$ marginals;
\item Unknown univariate margins, estimated non-parametrically and transformed to uniform scores. Copula parameters are estimated using the pseudo likelihood (\ref{loglik});
\item Unknown univariate margins, but well-specified marginal model. The margins have the Student-$t$ distribution with mean $m=1.5$, standard deviation $\sigma=0.85$ and $\nu = 8$ degrees of freedom. Marginal parameters are estimated first and then the integral transform is used to convert data to uniform marginals. The copula parameters are then estimated using the pseudo likelihood (\ref{loglik});
\item Same as in procedure 3 but all parameters are estimated jointly using (\ref{jloglik}).
\end{enumerate}

We first present results for two sets of parameters: $\tht = (1.2, 2.5, 1.2, 1.5)^{\TT}$ and $\tht = (0.8, 1.1, 0.7, 0.5)^{\TT}$ (with weak and strong dependence, respectively) for a $10\times10$ grid with $N=2000$ replicates; see Table \ref{tab-mse-expmle}. As expected, the procedure 1 gives the best results in terms of bias and standard deviation and provides an optimal benchmark for the other procedures. In more realistic scenarios, where margins are unknown and need to be estimated, we found that estimates obtained with procedure 2 are less accurate and can be quite heavily biased, especially with a larger number of locations $n$ or when dependence in the tails is weaker. Nevertheless, when the number of locations, $n$, is small relative to $N$ (e.g., $n=9$ and $N=2000$), procedure 2 yields quite accurate results. The estimates with procedure 4 are very accurate and the two-step estimation procedure 3 gives estimates with a higher standard deviation as this method is less efficient than estimating all parameters jointly in one step.

\begin{figure}[t!]
\begin{center}
\includegraphics[width=6.5in,height=1.7in]{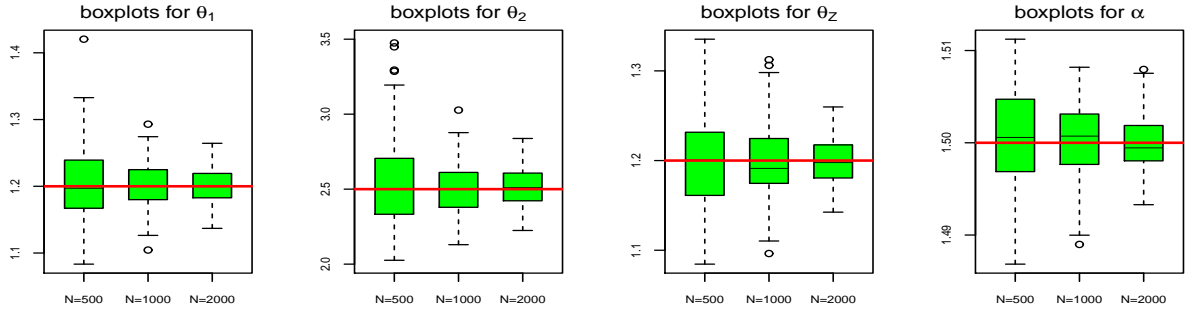}
\caption{{\footnotesize Boxplots of estimated copula parameters using procedure 4 (joint estimation of marginal and copula parameters) for data sets with $n=100$ locations and $N=500$ (left), $N=1000$ (middle) and $N=2000$ replicates (right); 500 experiments were performed. Red line shows true values of parameters.}}
\label{box1}
\end{center}
\end{figure}

\begin{figure}[]
\begin{center}
\includegraphics[width=6.5in,height=1.7in]{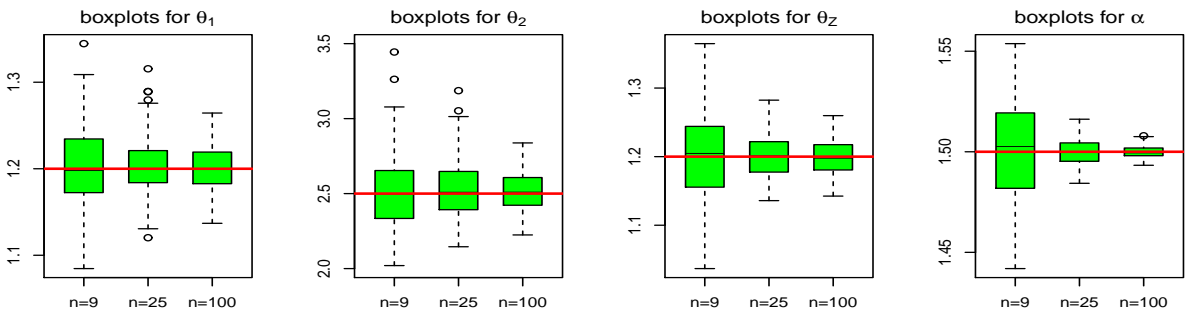}
\caption{{\footnotesize Boxplots of estimated copula parameters using procedure 4 (joint estimation of marginal and copula parameters) for data sets with $N=2000$ replicates and $n=9$ (left), $n=25$ (middle) and $n=100$ locations (right); 500 experiments were performed. Red line shows true values of parameters.}}
\label{box2}
\end{center}
\end{figure}

We now report more detailed results using procedure 4 with $\tht = (1.2, 2.5, 1.2, 1.5)^{\TT}$, where marginal and copula parameters are estimated jointly. Fig. \ref{box1} shows boxplots of the estimated copula parameters for data sets generated on a $10\times10$ grid with $N=500, 1000$ and $2000$ replicates. With a larger sample size, both the bias and variability of the maximum likelihood estimates decrease as expected because the maximum likelihood estimator is consistent as $N \to \infty$. Moreover, the width of boxplots decreases roughly proportionally to $\sqrt{N}$. Fig. \ref{box2} shows boxplots for data sets with $N = 2000$ replicates but with different numbers of spatial locations: 9, 25 and 100. One can see that the bias and standard deviation also decrease in this case although the maximum likelihood estimator is not necessarily consistent if $N$ is fixed, as $n\to \infty$. This is because the domain in our study is fixed $([0,1]\times[0,1])$ and many spatial estimators are inconsistent under infill asymptotics. Nevertheless, the maximum likelihood estimator has a much better performance for larger $n$, especially for the parameter $\alpha$, which controls the smoothness of the realized random field. Similar results were observed by \cite{Huser.Genton2016} among others.

\subsection{Simulation experiment under model misspecification} \label{sec_empstudy_sub1a}

In this section, we show that model (\ref{mainmodel}) with the exponential common factor can fit data well even if the distribution of $V_0$ is misspecified. We simulated data on a $5 \times 5$ uniform grid on $[0,1] \times [0,1]$ (so that there are $n=25$ locations) from the Pareto common factor model (\ref{mainmodel}) with $V_0 = V_1 - V_2$, where $V_1 \sim \mathrm{Pareto}(0.8,3)$, $V_2 \sim \mathrm{Pareto}(2.5,5)$; the powered-exponential correlation function, $\rho(h) = \exp(-0.6h^{1.2})$ was used to model the correlation structure of $Z$. We fitted the following misspecified models to the simulated data: 1) Factor copula model with $V_0 = V_1 - V_2$, where a) $V_j \sim \mathrm{Pareto}(\theta_j, 4)$; b) $V_j \sim \mathrm{Exp}(\theta_j)$ ($j=1,2$); and 2) Gaussian copula with no common factor.

To compare the fitted dependence structures with the data, we calculated the empirical and fitted model-based Spearman's correlation matrices. We denote these $n\times n$ matrices by $\widehat\rrho_S$ and $\widehat\rrho_S^{\mathrm{MLE}_m}$, respectively, where $m=1a,1b,2$ denotes the estimated model. However, the Spearman's correlation is not a good measure of dependence for the tails of a multivariate distribution. To compare the tail behavior of the two models, we therefore used the tail-weighted measures of dependence proposed by \citet{Krupskii.Joe2015}. 
The measures provide useful summaries of the strength of the tail dependence for each pair of variables, with values close to 0 or 1 corresponding to very weak (strong, respectively) dependence in the tails. Unlike many goodness-of-fit procedures studied in the literature, the tail-weighted measures of dependence give information on how the model can be improved to fit data better in the tails. Furthermore, they are more robust than tail dependence coefficients that can only be accurately estimated with a big sample size. We denote the empirical and fitted, model-based, tail-weighted measures of dependence in the lower/upper tail by $\widehat\aalpha_L/\widehat\aalpha_U$ and $\widehat\aalpha_L^{\mathrm{MLE}_m}/\widehat\aalpha_U^{\mathrm{MLE}_m}$ ($m=1\mbox{a},1\mbox{b},2$), respectively. We obtained the model-based estimates by simulating $10^5$ samples from models 1a, 1b and 2 with parameters set to the corresponding maximum likelihood estimates. We calculated
\begin{eqnarray*}
\Delta_{\rho,m} &:= & \frac{1}{n^2}{\sum}_{j_1,j_2=1}^n [\widehat\rrho_S-\widehat\rrho_S^{\mathrm{MLE}_m}]_{j_1,j_2},\,\,\, ~|\Delta_{\rho,m}| := \frac{1}{n^2}{\sum}_{j_1,j_2=1}^n |[\widehat\rrho_S-\widehat\rrho_S^{\mathrm{MLE}_m}]_{j_1,j_2}|,\\
\Delta_{L,m} &:= & \frac{1}{n^2}{\sum}_{j_1,j_2=1}^n [\widehat\aalpha_L-\widehat\aalpha_L^{\mathrm{MLE}_m}]_{j_1,j_2},\,\,\,  |\Delta_{L,m}| := \frac{1}{n^2}{\sum}_{j_1,j_2=1}^n |[\widehat\aalpha_L-\widehat\aalpha_L^{\mathrm{MLE}_m}]_{j_1,j_2}|,\\
\Delta_{U,m} &:= & \frac{1}{n^2}{\sum}_{j_1,j_2=1}^n [\widehat\aalpha_U-\widehat\aalpha_U^{\mathrm{MLE}_m}]_{j_1,j_2},\,\,\,  |\Delta_{U,m}| := \frac{1}{n^2}{\sum}_{j_1,j_2=1}^n |[\widehat\aalpha_U-\widehat\aalpha_U^{\mathrm{MLE}_m}]_{j_1,j_2}|.
\end{eqnarray*}
The results are reported in Table \ref{tab-gof-mle1}. They show that, although the scale parameter $\beta = 4$ for the Pareto factors in model 1a is different from the true value, the estimated structure fits the data quite well. Moreover, if the distribution for the common factor is misspecified as in model 1b, one can still get a model that fits the data reasonably well. Similar results are obtained with different choices of parameters. We therefore suggest using the exponential common factor model (Example 2 in Section \ref{sec_factmodel}) as a tractable, parsimonious, and fairly flexible model. Parameter estimation for this model is almost instantaneous, and the strength of dependence can be controlled in the lower and upper tails, depending on the choice of parameters $\theta_1$ and $\theta_2$. The Gaussian copula fails to handle tail dependence, as shown by model 2 significantly underestimating the strength of the lower tail dependence.

\subsection{Application to temperature data}
\label{sec_empstudy_sub2}

We fit our factor copula model (\ref{base_model}) to investigate the joint behavior of daily mean temperatures in Switzerland and compare the model's performance with some other popular models. We select 10 monitoring stations located in Switzerland: 1.~Basel-Binningen, 2.~Bern-Zollikofen, 3.~Buchs-Aarau, 4.~Cham,  5.~Fahy, 6.~Luzern, 7.~Neuch\^{a}tel, 8.~Payerne,  9.~Runenberg and 10.~Wynau. The  minimum and maximum altitudes of the selected stations are 316 and 611 meters, respectively. Therefore, we do not expect the altitude to have a significant effect on temperature for these stations. All stations are fairly close geographically, between the Alps and Jura mountains, and are typically subject to common weather patterns. The latter might thus be modelled as a ``common latent random factor'' affecting the region of study, hence providing support for our factor copula models. We use only the measurements obtained from May to September 2011, accounting for 153 days in total. Because of the short period of observations, we expect the marginal and the joint distributions of daily temperatures to be near-stationary.

For data with spatio-temporal dependence, one can specify a marginal distribution and estimate its parameters in a first step. The joint dependence of the residuals from the estimated marginal model (filtered data) can then be modeled by the factor copula model proposed in this paper. For the univariate marginals, we use an autoregressive-moving-average (ARMA) model to account for temporal dependence. In the summer time, there might be long periods of sunny weather with high temperatures, resulting in a strong temporal dependence. Also, the mean temperatures are higher in the middle of the selected time period. We therefore use a marginal model with one autoregressive lag and one moving average lag with a quadratic trend and skew-$t$ innovations \citep{Azzalini.Capitanio2003}: 
$$
M_{t,j} = \alpha_0 + \alpha_1 t + \alpha_2 t^2 + \beta_1 M_{t-1,j} + \epsilon_{t,j} + \gamma_1 \epsilon_{t-1,j}, \quad \epsilon_{t,j} \overset{\mathrm{i.i.d.}}{\sim} \text{Skew-}t(\nu, \delta),
$$
where $M_{t,j}$ is the mean temperature measured at the $j$-th station on day $t$ ($t=1,\ldots,153$). The parameters $\nu$, $\delta$ are the degrees of freedom and the skewness parameter, respectively. The parameters $\alpha_0, \alpha_1, \alpha_2, \beta_1, \gamma_1, \nu, \delta$ do not depend on the index $j$, so that we use the same marginal model for all stations. This is because we found that adding spatial covariates (latitude, longitude and altitude) did not significantly improve the fit.

We checked uncorrelatedness of the residuals using the Ljung-Box test. The p-value of the test at lags 1 to 20 is greater than 0.05 for all variables. The filtered data (residuals for each station $j$) were then transformed to the uniform scale using ranks, $u_{t,j} = \{\mathrm{rank}(\epsilon_{t,j} - 0.5)/153\}\  (t = 1,\ldots,153)$. 
The parametric integral transform could be used as well, but we found no significant difference between the two methods. In fact, with $n=10$ locations, a non-parametric approach works quite well as we mentioned in Section \ref{sec_empstudy_sub1b}.

We used the ranked data to compute the Spearman's $\rho$, and tail-weighted measures $\alpha_L$ and $\alpha_U$ for each pair of variables. We found that the values of $\alpha_L$ are mostly larger than those of $\alpha_U$, suggesting that dependence is stronger in the lower tail. We use normal score plots to visualize the dependence structure. To obtain normal scores, the ranked data are transformed to standard normal variables by applying the standard normal inverse distribution function: $z_{t,j} = \Phi^{-1}(u_{t,j})$ ($t=1,\ldots,153$, $j=1,\ldots,10$). If the residuals had a multivariate normal dependence structure, the normal scores would form ellipses. In the presence of tail dependence, the tails of the cloud of points are sharper.

\begin{figure}
\begin{center}
\includegraphics[width=6in,height=2in]{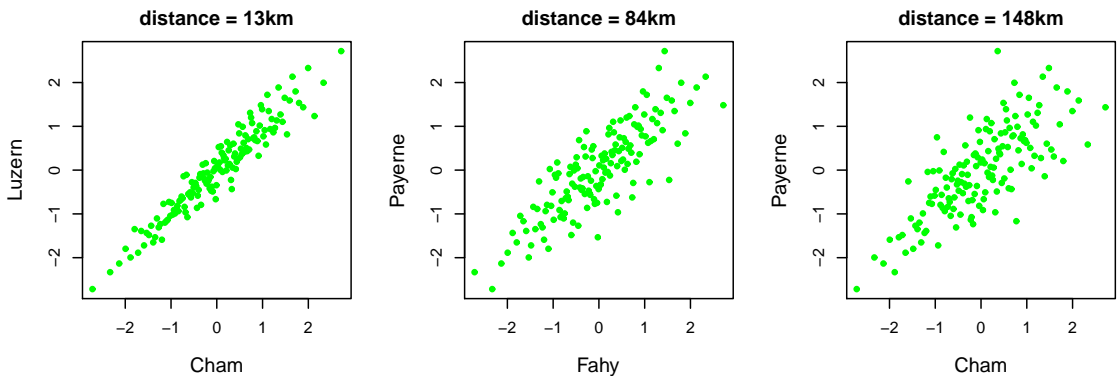}
\caption{{\footnotesize Scatter plots of normal scores for daily mean temperatures: Cham, Luzern (left); Fahy, Payerne (middle); Cham, Payerne (right).}}
\label{fig_sp_temp2011}
\end{center}
\end{figure}

In Fig. \ref{fig_sp_temp2011}, we see that the scatter plots of normal scores for these pairs of stations have sharp tails and that dependence in the lower tail is stronger than the upper tail. This implies that the data have both tail dependence and asymmetric dependence, suggesting that classical models based on the normal and Student's $t$ copulas might not be appropriate. Nevertheless, we include these two models for comparison. We fit the following models: 1) The normal copula; 2) The Student's $t$ copula; 3) The common factor model with $V_0 = V_1-V_2$, where $V_1, V_2$ are independent and a) $V_j\sim \mathrm{Pareto}(\theta_j,4)$; b) $V_j\sim \mathrm{Exp}(\theta_j)$ ($j=1,2$).

As mentioned in Section \ref{sec_empstudy_sub1a}, it is difficult to obtain accurate estimates for the scale and shape parameters for the Pareto common factor, and different parameters may result in models with very close dependence properties. We therefore fix the shape parameter for Pareto factors, $V_1$, $V_2$, in model 3a, and set it equal to 4. We do not use smaller values for the shape parameter as they result in very strong dependence, which may not be realistic. For all the above models we use the powered-exponential correlation function, $\rho(h) = \exp(-\theta_Z h^{\alpha})$, $\theta_Z > 0,\ 0< \alpha \leq 2$. For these models we calculate the maximum likelihood estimates and then compute $\Delta_{\rho,m}$, $|\Delta_{\rho,m}|$, $\Delta_{L,m}$, $|\Delta_{L,m}|$ and $\Delta_{U,m}$, $|\Delta_{U,m}|$, as defined in Section \ref{sec_empstudy_sub1a}, for models $m=1,2,3\mbox{a},3\mbox{b}$. The results are presented in Table \ref{tab-data-mle1}.

We can see that the covariance structure is well estimated for all models, though there are significant differences in the tails. The likelihood value is mostly influenced by the data in the middle of the distribution; therefore using Akaike/Bayesian information criteria (AIC/BIC) for model selection may be not appropriate as far as the fit in the tails is concerned. Indeed, both the normal and Student's $t$ copulas underestimate dependence in the lower tail and overestimate it in the upper tail; this is because these copulas do not allow for asymmetric dependence. However, the Student's $t$ copula yields a likelihood value fairly close to the one obtained in model 3b. The Pareto common factor model has a better fit in the tails and the exponential common factor model has the best fit in the tails. Overall, common factor models can handle both tail dependence and asymmetry, while fitting the data quite well in the center of the distribution.

\begin{figure}[b!]
\begin{center}
\includegraphics[width=6.4in,height=6.4in]{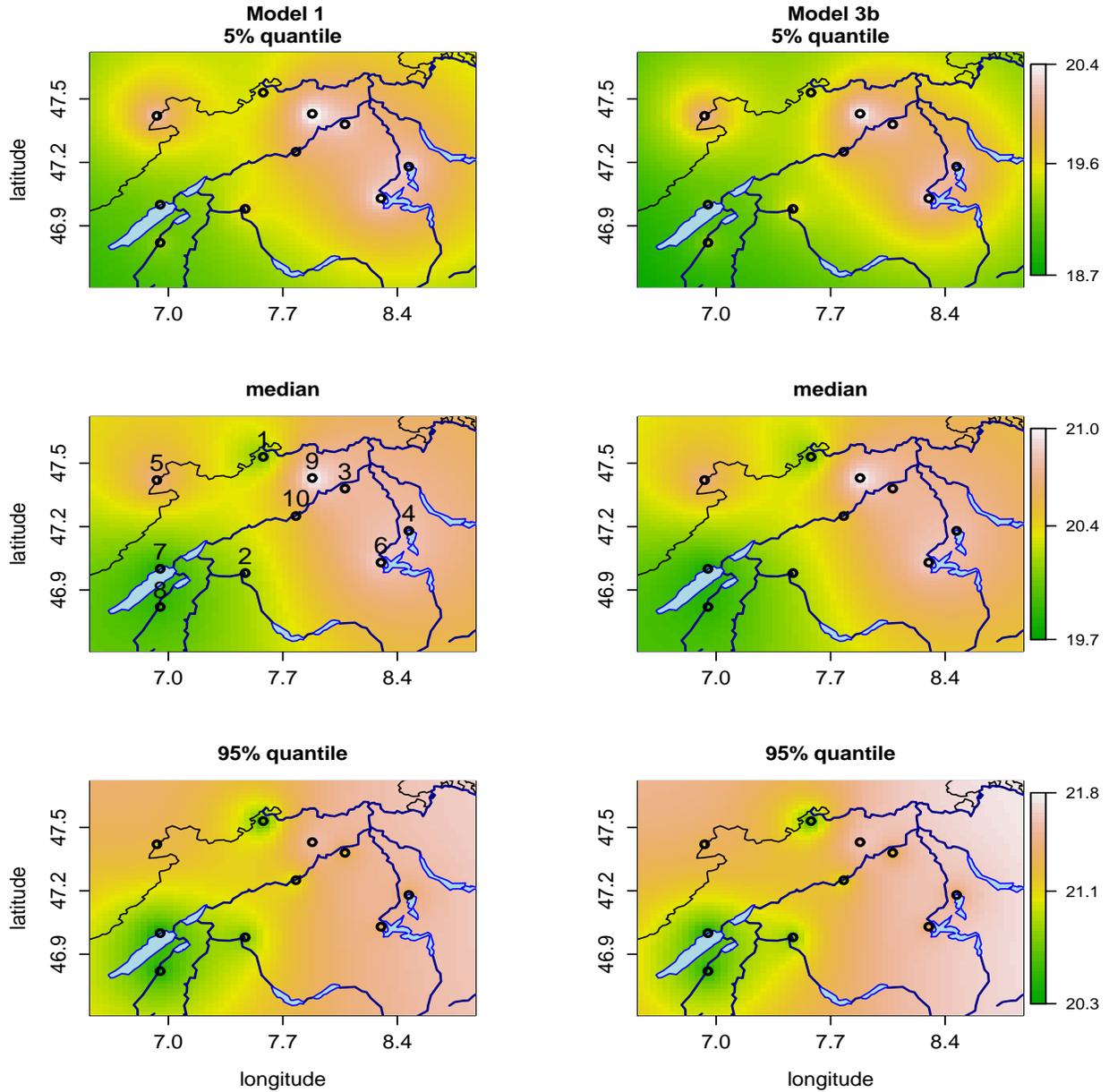}
\caption{{\footnotesize The predicted 5\% (top), 50\% (middle) and 95\% (bottom) quantiles for the Gaussian model 1 (left) and the new factor copula model 3b (right) for mean daily temperatures in the area of study (degrees Celsius), calculated for August 1, 2011. The 10 stations with recorded temperature data are shown as circles (numbers refer to station names in Section 4.3). The black line is the border of Switzerland and the blue lines are the main rivers.}}
\label{pred_mean_temp2011}
\end{center}
\end{figure}

Finally, we compute the predicted quantiles for the mean temperatures as shown in Section \ref{sec_MLE_sub3} for a $60 \times 60$ uniform grid in the region located between $46.6^{\circ}$ and $47.7^{\circ}$ North and between $6.5^{\circ}$ and $8.9^{\circ}$ East. We use the model with the best fit (model 3b) and the Gaussian copula (model 1) for comparison. We construct the map for the predicted medians as well as the 5\% and 95\% quantiles, conditional on the observed values on August 1, 2011; see Fig. \ref{pred_mean_temp2011}. One can clearly see the differences between these two models as expressed by lower 5\% and greater 95\% predicted quantiles for the exponential common factor model.

\section{Discussion}
\label{sec_discussion}

We proposed a new common factor copula model for spatial data. Unlike classical models in the literature, this large family of models can handle tail dependence and tail asymmetry. The common factor structure makes interpretation in practical applications easier than do vine copula models, in which the structure depends on the likelihood value and the vine construction. Maximum likelihood estimation can be quite easily performed using numerical integration. For some common factors, the joint density is available in closed form and therefore estimation is very fast even if the number of spatial locations is fairly large.

Despite its flexibility, the proposed model requires replicates for consistent inference. The reason is that the underlying spatial process (\ref{base_model}) is not ergodic, which entails large-scale dependence. One remedy might be to consider mixtures of truncated processes constructed using a compact random set \citep{Huser.Davison2014}. This construction would also allow to capture independence at large distances and therefore the resulting model might be applied to data in larger domains. Alternatively, a spatial model with a nested structure of independent random factors may be envisioned.

The proposed factor copula model can be naturally extended to a multivariate spatial process with $K$ variables measured at $n$ different locations. Define
$$
\mathrm{W}_{jk} = \mathrm{Z}_{jk} + V_{k} + V_{0}^* \quad (j = 1, \ldots, n, k = 1,\ldots,K),
$$
where $V_{k}, V_{0}^*$ are independent. Here, $V_k$ is a common factor for the $k$-th variable and $V_0^*$ is a common factor for all variables. The properties of this extended model, which depend on the choice of common factors, $V_k$ and $V_0^*$, will be a topic for future research. Another research direction is to include different types of common factors, for example models with multiplicative common factors \citep{Opitz2015}, which might be plausible in some applications, e.g., related to the modeling of extremes;  see \cite{Fereira.DeHaan2014}. 

\setcounter{section}{1}
\renewcommand{\thesection}{\Alph{section}}
\section*{Appendix}

\subsection*{Proof of Proposition \ref{prop-1}} \label{appx-prop1}

We have:
\begin{eqnarray*}
F_{2}^{\WW}(z,z) &:=& \pr(W_1 \leq z, W_2 \leq z) = \int_{\mathbb{R}^2}\pr(V_0 \leq z-w_1, V_0 \leq z-w_2)\phi_{\rho}(w_1,w_2)dw_1dw_2\\
&=& 2\int_{\mathbb{R}^1}\pr(V_0 \leq z-w_1)\int_{-\infty}^{w_1}\phi_{\rho}(w_1,w_2)\d w_2 \d w_1,
\end{eqnarray*}
and
\begin{eqnarray*}
\int_{-\infty}^{w_1}\phi_{\rho}(w_1,w_2)\d w_2 &=& \frac{\p\,\pr\{Z_1 \leq w, Z_2 \leq w_1\}}{\p w}\big|_{w=w_1}\\
&=& \pr\{Z_2 \leq w_1 | Z_1 = w_1\}\phi(w_1)  = \phi(w_1)\Phi\left\{\left(\frac{1-\rho}{1+\rho}\right)^{1/2}w_1\right\}.
\end{eqnarray*}
It implies
\begin{eqnarray*}
F_{2}^{\WW}(z,z) &=& 2\int_{\mathbb{R}^1}\pr(V_0 \leq z-w)\phi(w)\Phi\left\{\left(\frac{1-\rho}{1+\rho}\right)^{1/2}w\right\}\d w, \nonumber\\
F_{1}^{\WW}(z) &=& \int_{\mathbb{R}^1}\pr(V_0 \leq z-w)\phi(w)\d w. 
\end{eqnarray*}
Similarly, we can show that
\begin{eqnarray}
\bar F_{2}^{\WW}(z,z) &:=& \pr\{W_1 \geq z, W_2 \geq z\} = 2\int_{\mathbb{R}^1}\pr(V_0 \geq z-w)\phi(w)\Phi\left\{-\left(\frac{1-\rho}{1+\rho}\right)^{1/2}w\right\}\d w, \nonumber
\end{eqnarray}
\begin{eqnarray}
\bar F_{1}^{\WW}(z) &:=& \pr\{W_1 \geq z\} = \int_{\mathbb{R}^1}\pr(V_0 \geq z-w)\phi(w)\d w. \label{u.tail2}
\end{eqnarray}

For $0 < \epsilon < 1 - \alpha/2$ and $w^*(z) = z^{\alpha/2+\epsilon}$, from (\ref{u.tail2}) we get: $\bar F_{1}^{\WW}(z) = I_1(w,z) + g_1(z)$, $\bar F_{2}^{\WW}(z,z) = 2I_2(w,z) + 2g_2(z)$, where
\begin{eqnarray*}
I_1(w,z) &:=& \int_{|w|<w^*(z)}\pr(V_0 \geq z-w)\phi(w)\d w, \\
I_2(w,z) &:=& \int_{|w|<w^*(z)}\pr(V_0 \geq z-w)\phi(w)\Phi\left\{-\left(\frac{1-\rho}{1+\rho}\right)^{1/2}w\right\}\d w,
\end{eqnarray*}
where $0 \leq g_j(z) \leq 2\Phi\{-w^*(z)\}$, $j=1,2$.

As $z \to \infty$, $z-w \to \infty$ for $|w|<w^*(z)$ and $g_j(z)\exp(\theta z^{\alpha}) \to 0$, $j=1,2$,  and therefore
$$
\lambda_U  = \lim_{z\to \infty} \frac{\bar F_{2}^{\WW}(z,z)}{\bar F_{1}^{\WW}(z)} = \lim_{z\to \infty} \frac{2\int_{|w|<w^*(z)}\exp(\theta z^{\alpha})\pr(V_0 \geq z-w)\phi(w)\Phi\left\{-\left(\frac{1-\rho}{1+\rho}\right)^{1/2}w\right\}\d w}{\int_{|w|<w^*(z)}\exp(\theta z^{\alpha})\pr(V_0 \geq z-w)\phi(w)\d w}.
$$

For $|w|<w^*(z)$, we have: $\pr(V_0 \geq z-w) \sim_{z\to\infty} K(z-w)^{\beta}\exp\{-\theta(z-w)^{\alpha}\}$ and thus
$$
\exp(\theta z^{\alpha})\pr(V_0 \geq z-w) \sim_{z\to\infty} P(w,z),  \quad P(w,z) := K(z-w)^{\beta}\exp[\theta\{z^{\alpha} - (z-w)^{\alpha}\}].
$$

\emph{Case 1:} $\alpha = 0$, $\beta < 0$. We have:
$$
P\{-w^*(z),z\} \leq P(w,z) \leq P\{w^*(z),z\}, \quad |w| < w^*(z)
$$
and therefore
$$
\lim_{z\to \infty}\left[\frac{P\{-w^*(z),z\}}{P\{w^*(z),z\}}\cdot M^*\right] \leq \lambda_U \leq \lim_{z\to \infty}\left[\frac{P\{w^*(z),z\}}{P\{-w^*(z),z\}}\cdot M^*\right]\,,
$$
where
$$
\lim_{z\to \infty}\frac{P\{-w^*(z),z\}}{P\{w^*(z),z\}} = \lim_{z\to \infty}\left\{\frac{1+w^*(z)/z}{1-w^*(z)/z}\right\}^{\beta} = 1,
$$
$$
M^* : = \lim_{z\to\infty}\frac{2\int_{|w|<w^*(z)}\phi(w)\Phi\left\{-\left(\frac{1-\rho}{1+\rho}\right)^{1/2}w\right\}\d w}{\int_{|w|<w^*(z)}\phi(w)\d w} = 1,
$$
since the integrand in the numerator is the skew-normal random variable density \citep{Azzalini.Capitanio2003}. It implies that $\lambda_U = 1$.

\emph{Case 2:} $\alpha=1, \beta=0$. We have: $P(w,z) = K\exp(\theta w)$ and
$$
\lambda_U  = \frac{2\int_{-\infty}^{\infty}\exp(\theta w)\phi(w)\Phi\left\{-\left(\frac{1-\rho}{1+\rho}\right)^{1/2}w\right\}\d w}{\int_{-\infty}^{\infty}\exp(\theta w)\phi(w)\d w}.
$$
It is easy to see that the denominator equals $\exp(\theta^2/2)$. The numerator can be calculated by differentiating with respect to $\rho$, and it is equal to $\exp\left(\theta^2/2\right)\Phi\left[-\theta\{(1-\rho)/2\}^{1/2}\right]$. Therefore, $\lambda_U  = 2\Phi\left[-\theta\{(1-\rho)/2\}^{1/2}\right]$.

\emph{Case 3:} $\alpha=1$ and $\beta > 0$ (the proof for $\beta < 0$ is similar). It is easy to see that
$$
P\{w^*(z),z\} \leq P(w,z) \leq P\{-w^*(z),z\}, \quad |w| < w^*(z)
$$
and
$$
\lim_{z\to \infty}\left[\frac{P\{w^*(z),z\}}{P\{-w^*(z),z\}}\cdot M^{**}\right] \leq \lambda_U \leq \lim_{z\to \infty}\left[\frac{P\{-w^*(z),z\}}{P\{w^*(z),z\}}\cdot M^{**}\right]\,,
$$
where
$$
\lim_{z\to \infty}\frac{P\{w^*(z),z\}}{P\{-w^*(z),z\}} = \lim_{z\to \infty}\left[\frac{\exp(\theta w)\{1-w^*(z)/z\}}{\exp(\theta w)\{1+w^*(z)/z\}}\right]^{\beta} = 1,
$$
$$
M^{**} : = \lim_{z\to\infty}\frac{2\int_{|w|<w^*(z)}\exp(\theta w)\phi(w)\Phi\left\{-\left(\frac{1-\rho}{1+\rho}\right)^{1/2}w\right\}\d w}{\int_{|w|<w^*(z)}\exp(\theta w)\phi(w)\d w} = 2\Phi\left\{-\theta\left(\frac{1-\rho}{2}\right)^{1/2}\right\},
$$
and therefore $\lambda_U = 2\Phi\left[-\theta\left\{(1-\rho)/2\right\}^{1/2}\right]$.

\emph{Case 4:} $\alpha \neq 0$, $\alpha \neq 1$, $\beta > 0$ (the proof for $\beta < 0$ is similar). As $z\to \infty$ and $0 < \alpha <2$,
{\footnotesize
\begin{multline}
 \int_{|w|>w^*(z)}\exp\left(\alpha \theta wz^{\alpha-1}\right)\phi(w)\Phi\left\{-\left(\frac{1-\rho}{1+\rho}\right)^{1/2}w\right\}\d w \leq(2\pi)^{-1/2}\int_{|w|>w^*(z)}\exp\left(\alpha \theta wz^{\alpha-1}-\frac{w^2}{2}\right)\d w \\
 = \exp\left(\frac{\alpha^2\theta^2z^{2\alpha-2}}{2}\right)\left[\Phi\{-w^*(z)-\alpha \theta z^{\alpha-1}\}+\Phi\{-w^*(z)+\alpha \theta z^{\alpha-1}\}\right] \to 0. \label{prop-1-tails}
\end{multline} }

Define
\begin{eqnarray*}
M_1(z) &:=& 2K\int_{|w|<w^*(z)}\exp(\theta z^{\alpha})\pr(V_0 \geq z-w)\phi(w)\Phi\left\{-\left(\frac{1-\rho}{1+\rho}\right)^{1/2}w\right\}\d w,\\
M_2(z) &:=& K\int_{|w|<w^*(z)}\exp(\theta z^{\alpha})\pr(V_0 \geq z-w)\phi(w)\d w.
\end{eqnarray*}

For $\alpha <1$ we have as $z \to \infty$:
$$K\exp(\alpha^- \theta z^{\alpha^--1}w) < P(w,z) < K\exp(\alpha^+ \theta z^{\alpha^+-1}w), \quad 0 < \alpha^- < \alpha < \alpha^+ < 1,$$
if $|w|<w^*(z)$, and for $|w|>w^*(z)$, we use (\ref{prop-1-tails}) to get:
$$
M_1^*(z;\alpha^-) < M_1(z) < M_1^*(z;\alpha^+), \quad M_2^*(z;\alpha^-) < M_2(z) < M_2^*(z;\alpha^+),
$$
where
\begin{eqnarray*}
M_1^*(z;\alpha) &=& 2K\cdot\int_{\mathbb{R}^1}\exp(\alpha \theta z^{\alpha-1}w)\phi(w)\Phi\left\{-\left(\frac{1-\rho}{1+\rho}\right)^{1/2}w\right\}\d w\\
&=& 2K\cdot\exp\left\{\frac{\alpha^2\theta^2z^{2(\alpha-1)}}{2}\right\}\Phi\left\{-\alpha^2\theta^2z^{2(\alpha-1)}\left(\frac{1-\rho}{2}\right)^{1/2}\right\},
\end{eqnarray*}
\begin{eqnarray*}
M_2^*(z;\alpha) &=& K\cdot\int_{\mathbb{R}^1}\exp(\alpha \theta z^{\alpha-1}w)\phi(w)\d w = K\cdot\exp\left\{\frac{\alpha^2\theta^2z^{2(\alpha-1)}}{2}\right\},
\end{eqnarray*}
and therefore $\lambda_U = \lim_{z\to \infty}\frac{M_1(z)}{M_2(z)} = 1.$

For $1 < \alpha < 2$, $\alpha < \alpha^+ <2$, $|w| < w^*(z)$ and $z\to \infty$, $P(w,z) \leq K\exp(\alpha^+\theta z^{\alpha^+-1}w)$ and
\begin{multline*}
P(w,z) \geq K\exp\left\{\alpha\theta z^{\alpha-1}w - \frac{\alpha(\alpha-1)}{2}\cdot\frac{\theta z^{\alpha-2}w^2}{1-|w/z|}\right\} \\
\geq K\exp\left\{\alpha\theta z^{\alpha-1}w - \alpha(\alpha-1)\cdot\theta z^{\alpha-2}w^2\right\} \end{multline*}
for $z$ large enough so that $|w/z| < 1/2$.

This implies that, for $\alpha > 1$ and $z\to \infty$,
\begin{eqnarray*}
M_1(z) &\leq& 2K\cdot\int_{\mathbb{R}^1}\exp(\alpha^+ \theta z^{\alpha^+-1}w)\phi(w)\Phi\left\{-\left(\frac{1-\rho}{1+\rho}\right)^{1/2}w\right\}\d w\\
&=& 2K\exp\left\{\frac{[\alpha^+]^2 \theta^2 z^{2(\alpha^+-1)}}{2}\right\}\Phi\left\{-\alpha^+\theta z^{\alpha^+-1}\left(\frac{1-\rho}{2}\right)^{1/2}\right\},\\
M_2(z) &\geq& K\int_{|w|\leq w^*(z)}\exp\left[\alpha \theta z^{\alpha-1}w - \alpha(\alpha-1)\theta z^{\alpha-2}w^2 \right] \phi(w)\d w\\
&=& \frac{K}{c_{\alpha}^{1/2}}\exp\left(\frac{\alpha^2\theta^2z^{2(\alpha-1)}}{2c_{\alpha}}\right)\left[\Phi\left\{\frac{w^*(z)-\alpha\theta z^{\alpha-1}}{c_{\alpha}^{1/2}}\right\}-\Phi\left\{\frac{-w^*(z)-\alpha\theta z^{\alpha-1}}{c_{\alpha}^{1/2}}\right\}\right],
\end{eqnarray*}
where 
$c_{\alpha} = 1 + 2\alpha(\alpha-1)\theta z^{\alpha-2}/3$.
It follows that $\lambda_U = \lim_{z\to \infty} M_1(z)/M_2(z) = 0$.
$\hfill \Box$

\baselineskip=16.9pt
\bibliographystyle{model2-names}
\bibliography{fmcop}

\newpage

\begin{table}
\caption{{\footnotesize Bias and standard deviation for maximum likelihood estimates in the exponential common factor model for 4 different procedures; standard errors are shown in small font. 500 experiments were done.}}
\label{tab-mse-expmle}
\begin{center}
\begin{tabular}{lcc}
&$\tht=(1.2, 2.5, 1.2, 1.5)^{\TT}$ & $\tht=(0.8, 1.1, 0.7, 0.5)^{\TT}$ \\
Procedure 1 & $(0.00_{0.01}, 0.00_{0.06},  ~~0.00_{0.02}, ~~0.00_{0.00})^{\TT}$&$(0.00_{0.01}, 0.00_{0.01}, ~~0.00_{0.01}, ~~0.00_{0.00})^{\TT}$ \\
Procedure 2 & $(0.25_{0.15}, 4.89_{4.76}, -0.20_{0.07}, -0.01_{0.00})^{\TT}$&$(0.06_{0.13}, 0.03_{0.07}, -0.03_{0.11}, -0.01_{0.01})^{\TT}$  \\
Procedure 3 & $(0.03_{0.08}, 0.36_{0.83}, -0.02_{0.08}, ~~0.00_{0.00})^{\TT}$&$(0.00_{0.11}, 0.00_{0.06}, ~~0.01_{0.11}, ~~0.01_{0.01})^{\TT}$ \\
Procedure 4 & $(0.00_{0.03}, 0.01_{0.13}, ~~0.00_{0.03}, ~~0.00_{0.00})^{\TT}$&$(0.00_{0.03}, 0.00_{0.02}, ~~0.00_{0.03}, ~~0.00_{0.00})^{\TT}$ \\
\end{tabular}
\end{center}
\end{table}

\begin{table}
\def~{\hphantom{0}}
\caption{{\footnotesize $\Delta_{\rho,m}, |\Delta_{\rho,m}|, \Delta_{L,m}, |\Delta_{L,m}|, \Delta_{U,m}, |\Delta_{U,m}|$ for different models $m$. Simulated data were used to calculate these values; number of replicates was $N = 10^5$. }}
\label{tab-gof-mle1}
\begin{center}
\begin{tabular}{lccc}
\\
~Model $m$ & \multicolumn{1}{c}{$\Delta_{\rho,m}/ |\Delta_{\rho,m}|$} & \multicolumn{1}{c}{$\Delta_{L,m}/ |\Delta_{L,m}|$} & \multicolumn{1}{c}{$\Delta_{U,m}/ |\Delta_{U,m}|$}  \\
~ 1a~ & ~~0.00/0.01~ & ~0.05/0.05~ & ~~0.01/0.03~ \\
~ 1b~ & ~$-$0.02/0.02~ & ~0.00/0.01~ & ~$-$0.01/0.02~ \\
~ 2~~ & ~$-$0.01/0.02~ & ~0.17/0.17~ & ~~0.02/0.04~ \\
\end{tabular}
\end{center}
\end{table}

\begin{table}
\def~{\hphantom{0}}
\caption{{\footnotesize $\Delta_{\rho,m}, |\Delta_{\rho,m}|, \Delta_{L,m}, |\Delta_{L,m}|, \Delta_{U,m}, |\Delta_{U,m}|$ for different models $m$. Simulated data were used to calculate these values; number of replicates was $N = 10^5$. }}
\label{tab-data-mle1}
\begin{center}
\begin{tabular}{lcccc}
\\
Model $m$ & Log-likelihood & \multicolumn{1}{c}{$\Delta_{\rho,m}/ |\Delta_{\rho,m}|$} & \multicolumn{1}{c}{$\Delta_{L,m}/ |\Delta_{L,m}|$} & \multicolumn{1}{c}{$\Delta_{U,m}/ |\Delta_{U,m}|$}  \\
\\
1~ & 1,342 & 0.00/0.03 & 0.13/0.13 & $-$0.10/0.12 \\
2~ & 1,369 & 0.00/0.03 & 0.11/0.11 & $-$0.13/0.14 \\
3a & 1,359 & 0.00/0.03 & 0.07/0.07 & $-$0.09/0.11 \\
3b & 1,381 & 0.00/0.03 & 0.02/0.04 & $-$0.03/0.09 \\
\end{tabular}
\end{center}
\end{table}

\end{document}